\documentclass[english]{article}
\usepackage[T1]{fontenc}
\usepackage[utf8]{inputenc}
\usepackage[a4paper]{geometry}
\usepackage{babel}
\usepackage{textcomp}
\usepackage[unicode=true]
 {hyperref}

 \usepackage{caption}
 \usepackage{subcaption}

 \usepackage{amsmath}
\usepackage{graphicx}
\usepackage{svg}
 \usepackage{tikz}
\usetikzlibrary{positioning}
\usetikzlibrary{backgrounds,scopes}

\usepackage{tipa}

\usepackage{lineno}

\usepackage[sorting=none, giveninits=true, maxbibnames=99]{biblatex}
\bibliography{zoterobib.bib}
\AtEveryBibitem{
\ifentrytype{article}{%
    \clearfield{url}%
    \clearfield{doi}%
    }{}%
\ifentrytype{inproceedings}{%
    \clearfield{url}%
    \clearfield{doi}%
    }{}%
\ifentrytype{book}{%
    \clearfield{url}%
    \clearfield{doi}%
    }{}%
\ifentrytype{incollection}{%
    \clearfield{url}%
    \clearfield{doi}%
    }{}%
    \clearlist{language} 
    \clearfield{urlyear} 
    \clearfield{note}%
    \clearfield{issn}%
    \clearfield{isbn}%
    \clearfield{eprint}%
}  
\usepackage{csquotes}

\makeatletter
\newcommand{\lyxaddress}[1]{
	\par {\raggedright 
 #1
	\vspace{1.4em}
	\noindent
 \par}
}

\@ifundefined{date}{}{\date{}}
\makeatother

\begin{document}
\title{Modeling of Speech-dependent Own Voice Transfer Characteristics for Hearables with an In-ear Microphone}
\author{Mattes Ohlenbusch$^{1*}$, Christian Rollwage$^1$, Simon Doclo$^{1,2}$}
\maketitle

\lyxaddress{$^{1}$ Fraunhofer Institute for Digital Media Technology IDMT,
Oldenburg Branch for Hearing, Speech and Audio Technology HSA, Germany}

\lyxaddress{$^{2}$ Carl von Ossietzky Universität Oldenburg, Department of Medical Physics and Acoustics and Cluster of Excellence Hearing4all, Germany}

\lyxaddress{$^{*}$ Corresponding author; e-mail: mattes.ohlenbusch@idmt.fraunhofer.de}
\begin{abstract}
\noindent Many hearables contain an in-ear microphone, which may be used to capture the own voice of its user.
However, due to the hearable occluding the ear canal, the in-ear microphone mostly records body-conducted speech, typically suffering from band-limitation effects and amplification at low frequencies. 
Since the occlusion effect is determined by the ratio between the air-conducted and body-conducted components of own voice, the own voice transfer characteristics between the outer face of the hearable and the in-ear microphone depend on the speech content and the individual talker. 
In this paper, we propose a speech-dependent model of the own voice transfer characteristics based on phoneme recognition, assuming a linear time-invariant relative transfer function for each phoneme.
We consider both individual models as well as models averaged over several talkers.
Experimental results based on recordings with a prototype hearable show that the proposed speech-dependent model enables to simulate in-ear signals more accurately than a speech-independent model in terms of technical measures, especially under utterance mismatch and talker mismatch.
Additionally, simulation results show that talker-averaged models generalize better to different talkers than individual models. 
\end{abstract} 

\clearpage

\section{Introduction}
\label{sec:introduction}
Hearables, i.e. smart earpieces containing a loudspeaker and one or more microphones, are often used for speech communication in noisy acoustic environments.
In this paper, we consider the scenario where the hearable is used to pick up the own voice of the user talking in a noisy environment (e.g., to be transmitted via a wireless link to a mobile phone or another hearable).
Assuming that the hearable is at least partly occluding the ear canal, in this scenario an in-ear microphone may be beneficial to pick up the own voice since environmental noise is attenuated. 
Compared to own voice recorded at the outer face of the hearable, own voice recorded inside an occluded ear is known to suffer from amplification at low frequencies (below ca.\,1\,kHz) and strong attenuation at higher frequencies (above ca.\,2\,kHz), leading to a limited bandwidth~\cite{bouserhal_-ear_2019}. 
The occlusion effect is determined by the ratio between the air-conducted and body-conducted components of own voice, which depends on device properties such as earmould fit and insertion depth~\cite{hansen_occlusion_1998-1},
individual anatomic factors such as residual ear canal volume and shape~\cite{stenfelt_model_2007,vogl_individualized_2019}, and the generated sounds or phonemes~\cite{reinfeldt_hearing_2010, saint-gaudens_towards_2022}.
In particular, it has been shown that the occlusion effect for different vowels can be predicted by a linear combination of their formant frequencies~\cite{zurbrugg_investigations_2014}, with closed front vowels exhibiting the largest occlusion effect.   
In addition, mouth movements during articulation~\cite{richard_effect_2023} and body-conduction from different places of excitation~\cite{porschmann_influences_2000}
likely influence the occlusion effect as well. 
Unlike acoustical models based on ear canal geometry~\cite{stenfelt_model_2007} or three-dimensional finite element models of body-conduction occlusion~\cite{brummund_three-dimensional_2014}, 
in this paper we consider a signal processing-based approach to model the own voice transfer characteristics between a microphone
at the entrance of the occluded ear canal (i.e. at the outer face of the hearable) and an in-ear microphone. 

In many hearable applications, acoustic transfer path models for the microphone inside the occluded ear canal are required.
For example, active noise cancellation algorithms may benefit from an accurate estimate of the so-called secondary path between the hearable loudspeaker and the in-ear microphone~\cite{liebich_signal_2018, rivera_benois_optimization_2022}.
In active occlusion cancellation (AOC), models of the own voice transfer path between the microphones inside and outside of the occluded ear canal can be used to generate a cancellation signal that aims at compensating the occlusion effect as measured at the in-ear microphone~\cite{zurbrugg_occlusion_2018, liebich_occlusion_2022}. 
Models of the own voice transfer path are not only relevant for AOC, but also for algorithms to enhance the quality of the in-ear microphone signal picking up the own voice of the user.
Several own voice reconstruction algorithms aiming at bandwidth extension, equalization and noise reduction have been proposed, e.g., based on classical signal processing~\cite{bouserhal_-ear_2017} or supervised learning~\cite{wang_fusing_2022, ohlenbusch_training_2022, hauret_configurable_2023, ohlenbusch_multi-microphone_2024}.
Supervised learning-based approaches typically require large amounts of training data. 
Since large amounts of realistic in-ear recordings may be hard to obtain for several talkers, an accurate and possibly individual model of the own voice transfer characteristics would be highly beneficial.
Such a model would enable to generate large amounts of simulated in-ear signals either from recordings at the entrance of the ear canal or from speech corpora, e.g.,~\cite{panayotov_librispeech_2015}. 
Data augmentation can then be performed with these simulated in-ear signals to train supervised learning-based own voice reconstruction algorithms.
Similarly as for other acoustic signal processing applications~\cite{ko_study_2017, he_neural_2021, srivastava_realistic_2022}, it is expected that using more accurate acoustic models for generating augmented training data improves system performance and generalization ability.

Several models of own voice transfer characteristics have been presented in the literature, either between two air-conduction microphones~\cite{ohlenbusch_training_2022} or between an air-conduction and a body-conduction microphone~\cite{wang_fusing_2022, hauret_configurable_2023, pucher_conversion_2021}. 
In~\cite{pucher_conversion_2021}, it has been proposed to convert air-conducted to bone-conducted speech using a deep neural network (DNN) model that accounts for individual differences between talkers based on a speaker identification system.
In~\cite{wang_fusing_2022}, a DNN model estimating bone-conducted speech from air-conducted speech is jointly trained with a multi-modal enhancement network within a semi-supervised training scheme, resulting in reduced data requirements compared to fully supervised training. 
Instead of using rather complicated black-box DNN models, in~\cite{ohlenbusch_training_2022, hauret_configurable_2023} time-invariant linear relative transfer functions (RTFs) are used to model own voice transfer characteristics.
To introduce variations in the simulated own voice signals, either RTFs estimated on recordings of multiple talkers are used~\cite{ohlenbusch_training_2022}, or random values are added to the magnitude of the RTF estimated from a single talker~\cite{hauret_configurable_2023}. 
It should be realized that these variations do not account for the speech-dependent nature of the own voice transfer characteristics.

Aiming at obtaining a model of the own voice transfer characteristics that generalizes well to unseen utterances and talkers, in this paper we propose a speech-dependent system identification approach, where for each phoneme a different RTF between the microphone at the entrance of the occluded ear canal and the in-ear microphone is estimated.
We consider both individual as well as talker-averaged models.
To simulate in-ear own voice signals from broadband speech, a phoneme recognition system is first utilized to segment the broadband speech into different segments corresponding to a specific phoneme, which are then filtered using the corresponding (smoothed) phoneme-specific RTFs.
In contrast to previous RTF-based modeling approaches~\cite{ohlenbusch_training_2022, hauret_configurable_2023}, the proposed model of own voice transfer characteristics is speech-dependent and thus time-varying.
In addition, contrary to the DNN-based modeling approach~\cite{wang_fusing_2022}, only a small amount of own voice recordings are required for model estimation.
The accuracy of simulating in-ear signals is assessed using recorded own voice signals of over 300 utterances by 18 talkers, each wearing a prototype hearable device~\cite{denk_one-size-fits-all_2019}.
The role of speech-dependency for simulating in-ear own voice signals is investigated by comparing the proposed speech-dependent RTF-based model to a speech-independent RTF-based model, and an adaptive filtering-based model~\cite{haykin_adaptive_1996} which is utterance-specific.
Experimental results show that the proposed speech-dependent model enables to simulate in-ear own voice signals more accurately than the speech-independent model and the adaptive filtering-based model in terms of technical distance measures. 
In addition, the performance of individual and talker-averaged models is compared in terms of their generalization capability to unseen talkers. 
Results show that the speech-dependent talker-averaged model generalizes better to utterances of unseen talkers compared to speech-independent or individual models.
Preliminary results of the proposed approach have already been published in~\cite{ohlenbusch_speech-dependent_2023}. 
This paper extends upon previous work presented in~\cite{ohlenbusch_speech-dependent_2023} by proposing talker-averaged models, by investigating utterance and talker mismatch separately, and by conducting experiments on a larger corpus of hearable recordings.

The paper is structured as follows: In Section~\ref{sec:sigmodel}, the own voice signal model is introduced. 
In Section~\ref{sec:models}, several system identification approaches to model own voice transfer characteristics using time-invariant or time-varying linear filters are presented. 
In Section~\ref{sec:evaluation}, the performance of these models is evaluated using recorded own voice signals for different conditions.

\section{Signal model}
\label{sec:sigmodel}
\begin{figure} 
    \center
    \begin{tikzpicture}
        \node[opacity=.4] (tepear) at (0,0) 
        {\includegraphics[width=5cm]
        {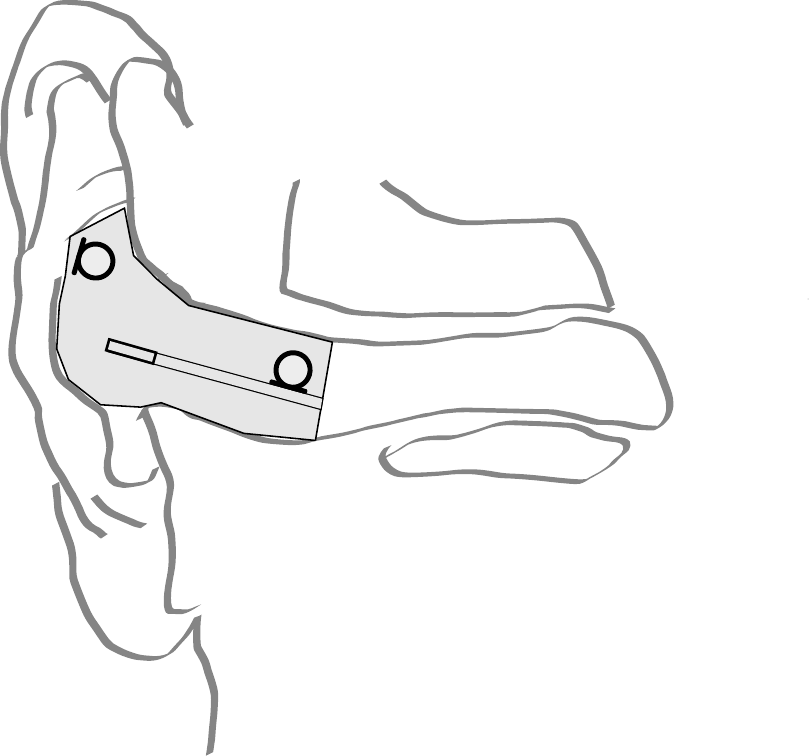}
        };
        \node[label={$y_o^a$}] (oem) at (-1.925,0.725) { };
        \node[label={$y_i^a$}] (iem) at (-0.6925,0.05) { };
        
        \node[below=of oem] (speech) {$s_o^a$};
        \node[below=1.5 of iem] (iemspeech) {$s_i^a$};
        \node[below right= of iem] (bodnoise) {$v_i^a$};
        \node[below left= of oem] (oemnoise) {$v_o^a$};

        \draw[->,very thick] (speech) -- (oem);
        \draw[->,very thick] (iemspeech) -- (iem);
        \draw[->, very thick] (bodnoise) -- (iem);
        \draw[->, very thick] (oemnoise) -- (oem);
        \draw[->, dashed, color=red,very thick] (speech) |-  node[draw,solid,fill=white] {$T^{a}$} (iemspeech);
        \vspace{-2cm}
    \end{tikzpicture}
    \caption{The own voice signal model for a hearable with two microphone (outer face, in-ear).} 
    \label{fig:sigmodel}
\end{figure}
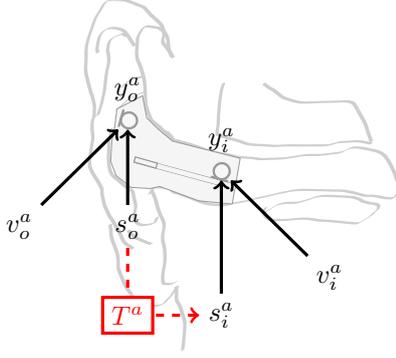
Figure~\ref{fig:sigmodel} depicts a hearable device equipped with an in-ear microphone and a microphone at the entrance of the (partly) occluded ear canal.
The signals at both microphones are denoted by subscripts $i$ and $o$, respectively. 
We assume that the hearable is worn by a person (referred to as talker) in a noiseless environment.
In the time domain, $s_i^a[n]$ and $s_o^a[n]$ denote the own voice component of talker $a$ at both microphones, where $n$ denotes the discrete-time index. 
The in-ear microphone signal $y_i^a[n]$ consists of the own voice component and additive noise, i.e.
\begin{equation}
    y_i^a[n] = s_i^a[n] + v_i^a[n],
\end{equation}
where the noise component $v_i^a[n]$ consists of unavoidable body-produced noise (e.g., breathing sounds, heartbeats).
Similarly, the microphone signal at the entrance of the occluded ear canal $y_o^a[n]$ can be written as
\begin{equation}
    y_o^a[n] = s_o^a[n] + v_o^a[n],
\end{equation}
where $v_o^a[n]$ mainly consists of sensor noise.
The sensor noise is assumed to be negligible compared to the own voice component in both microphone signals.
The own voice components of talker $a$ at the in-ear microphone and the microphone at the entrance of the occluded ear canal $s_\mathrm{o}^a[n]$ are assumed to be related by the own voice transfer characteristics $T^a\{\cdot\}$, i.e.
\begin{equation}
    s_\mathrm{i}^a[n] = T^a\left\{s_\mathrm{o}^a[n]\right\}.
\end{equation}
Due to individual anatomical differences of the ear canal~\cite{vogl_individualized_2019}, these transfer characteristics depend on the talker.
In addition, it has been shown that these transfer characteristics depend on the spoken sounds~\cite{reinfeldt_hearing_2010,saint-gaudens_towards_2022} (see also Figure~\ref{fig:rtfs_timedomain_annotation_phones}).

In this paper, we assume that the own voice transfer characteristics $T^a\{\cdot\}$ can be modeled as a \textit{time-varying linear system}, i.e.
\begin{equation}
    s_i^a[n] = H^a(q,n) \cdot s_o^a[n],
    \label{eq:transfer_characteristics_q}
\end{equation}
with 
\begin{equation}
    H^a(q,n) = \mathbf{h}^T[n] \mathbf{q}.
\end{equation}
The vector $\mathbf{h}[n]$ denotes a time-varying finite impulse response (FIR) filter with $N$ coefficients, 
\begin{equation}
    \mathbf{h}[n] = \begin{bmatrix} h_0[n], & h_1[n] & \dots, & h_{N-1}[n] \end{bmatrix}^T, 
\end{equation}
with $\{\cdot\}^T$ the transpose operator, and the vector $q$ is defined as~\cite{ljung_system_1998} 
\begin{equation}
    \mathbf{q} = \begin{bmatrix} 1, & q^{-1}, & \dots, & q^{-N+1} \end{bmatrix}^T,
\end{equation}
with $q^{-1}$ the delay operator.
The filtering operation in~\eqref{eq:transfer_characteristics_q} can be approximated in the short-time Fourier transform (STFT) domain as
\begin{equation}
    S_\mathrm{i}^a(k,l) = H^a(k,l) \cdot S_\mathrm{o}^a(k,l),
\end{equation}
where $k$ denotes the frequency bin index, $l$ denotes the time frame index and $H^a(k,l)$ denotes the relative transfer function (RTF) between the microphone at the entrance of the occluded ear canal and the in-ear microphone. 
Different from \eqref{eq:transfer_characteristics_q}, this approximation is only time-varying between STFT frames and not within a single STFT frame\footnote{Circular convolutions effects are also neglected in this approximation, but can be reduced by appropriate windowing.}.

\section{Modeling of own voice transfer characteristics}
\label{sec:models}
\begin{figure}
\center
 \includegraphics
 {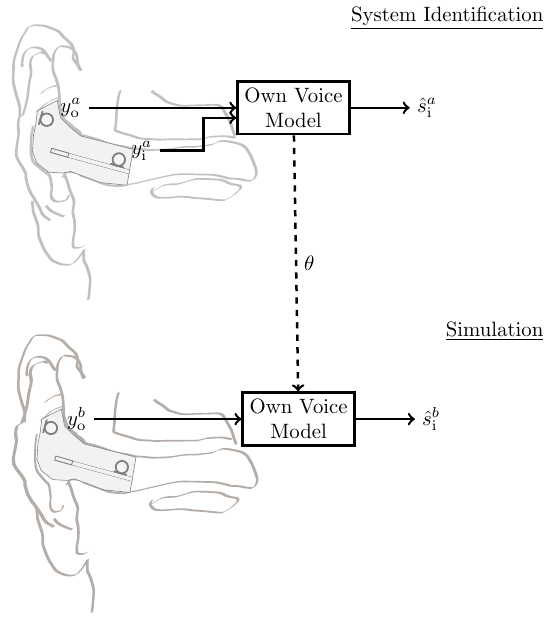} 
 \caption{Overview of the identification and simulation steps of the own voice transfer characteristic models.}
 \label{fig:simulation_diagram}
\end{figure}
In this section, several methods are presented to model own voice transfer characteristics and subsequently simulated in-ear own voice signals.
As outlined in Fig.~\ref{fig:simulation_diagram}, in the \textit{system identification step} the parameters $\theta$ of the model $\hat{T}_\theta\{\cdot\}$ are estimated (either in time domain or in frequency domain) based on the signals recorded at the in-ear microphone and the microphone at the entrance of the occluded ear canal. 
In the \textit{simulation step}, this model can then be used to generate simulated in-ear own voice signals from microphone signals at the entrance of the occluded ear canal, i.e.
\begin{equation}
    \hat{s}_i^b[n] = \hat{T}_\theta\left\{y_o^b[n]\right\}.
\end{equation}
Both individual models for a specific talker as well as talker-averaged models will be considered. 
In Section~\ref{sec:evaluation} it will be experimentally investigated whether talker-averaging increases robustness to talker mismatch.
To estimate the individual model $\hat{T}_\theta^a$ for talker $a$, recorded microphone signals from talker $a$ are used. 
This model can then be used to simulate in-ear signals either for the same talker $a$ and the same recorded microphone signals (same talker, same utterance), for different utterances of talker $a$ than used during system identification (utterance mismatch), or for utterances of another talker $b$ (talker mismatch).
To estimate the talker-averaged model $\hat{T}_\theta^\text{avg}$, recorded microphone signals from several talkers are used.

Sections~\ref{sec:speechindepmodel}-\ref{sec:avgmodels} consider RTF-based frequency-domain models for the own voice transfer characteristics.
In Section~\ref{sec:speechindepmodel}, a speech-independent time-invariant model for a specific talker is presented, similarly as in~\cite{ohlenbusch_training_2022}.
In Section~\ref{sec:speechdepmodel}, a speech-dependent model for a specific talker is proposed,  which accounts for the time-varying own voice transfer characteristics by assuming a different RTF for each phoneme. 
Section~\ref{sec:avgmodels} describes how to compute talker-averaged speech-independent and speech-dependent models.
Contrary to Sections~\ref{sec:speechindepmodel}-\ref{sec:avgmodels}, in Section~\ref{sec:adaptmodel} an adaptive filtering-based time-domain model of own voice transfer characteristics is presented, which is utterance-specific.

\subsection{Speech-independent individual model}
\label{sec:speechindepmodel}
\begin{figure}
    \centering
    \includegraphics[page=5]{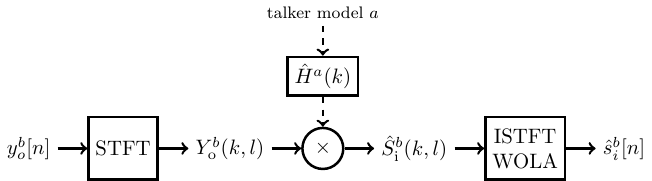} 
    \caption{Simulation of in-ear own voice signals for talker $b$ using the speech-independent model for talker $a$.}
    \label{fig:signalflow_sim_fix_indiv} 
\end{figure}

If own voice transfer characteristics are assumed to be speech-independent, the individual transfer characteristics of talker $a$ can be modeled as a time-invariant RTF $H^a(k)$ between the  microphone at the entrance of the occluded ear canal and the in-ear microphone:
\begin{equation}
    \theta_\mathrm{sp.-indep.}^a = \left\{\hat{H}^a(k)~\middle|~k=1,~\dots,~K\right\},
\end{equation}
where $K$ denotes the STFT size.
Assuming that the own voice component $S_o^a$ at the entrance of the occluded ear canal and the body-produced noise $V_i^a$ are independent, in the \textit{system identification step} the RTF $\hat{H}^a(k)$ can be estimated using the well-known least squares approach~\cite{avargel_multiplicative_2007}, i.e.
\begin{equation}
\hat{H}^a(k) = \arg\min_{H^a(k)} \sum_{l} |Y_\mathrm{i}^a(k,l) - H^a(k) \cdot Y_\mathrm{o}^a(k,l)|^2,
\end{equation}
considering all STFT frames of the recorded microphone signals from talker $a$ used for system identification. 
The least-squares RTF estimate is obtained as 
\begin{equation}
\hat{H}^a(k) = \frac{ \sum_l Y_\mathrm{i}^a(k,l) \cdot Y_\mathrm{o}^{a,*}(k,l) }{ \sum_l |Y_\mathrm{o}^a(k,l)|^2},
\end{equation}
where $\cdot^*$ denotes complex conjugation.
In the \textit{simulation step}, own voice speech of talker $b$ recorded at the microphone at the entrance of the occluded ear canal is filtered in the STFT domain with the RTF estimate of talker $a$ (where talker $a$ and $b$ can be the same or different), i.e.
    \begin{equation}
        \hat{S}_\mathrm{i}^b(k,l) = \hat{H}^a(k) \cdot Y_\mathrm{o}^b(k,l).
    \label{eq:simulation_speechindep}
    \end{equation}
After applying the inverse STFT, a weighted overlap-add (WOLA) scheme is employed to obtain the time domain signal $\hat{s}_i^b[n]$.
Figure~\ref{fig:signalflow_sim_fix_indiv} depicts the signal flow to simulate in-ear own voice signals for talker $b$ using the speech-independent individual model for talker $a$.

\subsection{Speech-dependent individual model} 
\label{sec:speechdepmodel}
Since own voice transfer characteristics likely depend on speech content, we propose to model the transfer characteristics $T^a$ of talker $a$ using a time-varying speech-dependent model. 
In the \textit{system identification step}, first a frame-wise phoneme annotation $p(l) \in 1,~\dots,~P$ with $P$ possible phoneme classes is obtained from the microphone signal $y_\mathrm{o}^a[n]$ at the entrance of the occluded ear canal using a phoneme recognition system $R\{\cdot\}$:  
\begin{equation}
        p(l) = R\left\{ y_\mathrm{o}^a[n] \right\}.
\end{equation}
Assuming that the transfer characteristics for each phoneme can be modeled using a (time-invariant) RTF, the RTF for phoneme $p^\prime$ can be estimated from all frames where this phoneme is detected as
\begin{equation}
    \hat{H}_{p^\prime}^a(k) = \frac
    {\sum_{p(l)=p^\prime}  Y_\mathrm{i}^a(k,l) \cdot Y_\mathrm{o}^{a,*}(k,l)}
    {\sum_{p(l)=p^\prime}  |Y_\mathrm{o}^a(k,l)|^2}.
    \label{eq:speechdep_indiv_estimation}
\end{equation}
Hence, the speech-dependent model for talker $a$ consists of $P$ RTFs:
\begin{equation}
    \theta_\mathrm{sp.-dep.}^a = \left\{ \hat{H}_{p}^a(k)~\middle|~p \in 1,\ \dots,\ P,\ k=1,\ \dots,\ K \right\}.
\end{equation}

\begin{figure}
    \centering
    \includegraphics[page=6]{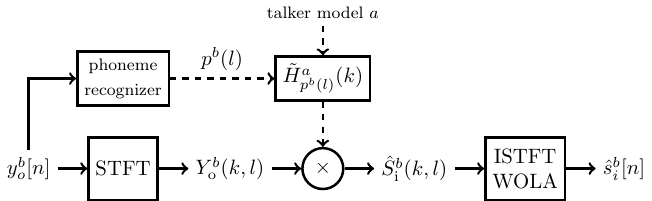} 
    \caption{Simulation of in-ear own voice signals for talker $b$ using the proposed speech-dependent model for talker $a$.}
    \label{fig:signalflow_sim_phone_indiv}
\end{figure}

In the \textit{simulation step}, first the phoneme sequence $p^b(l)$ is determined on the own voice speech of talker $b$ recorded at the microphone at the entrance of the occluded ear canal.
For each frame, the corresponding phoneme-specific RTF $\hat{H}_{p^b(l)}^a(k)$ is selected.
In order to prevent discontinuities in the RTFs during phoneme transitions, recursive smoothing with smoothing constant $\alpha$ is applied, i.e. 
\begin{equation}
    \tilde{H}_{p^b(l)}^a(k) = \alpha \cdot \tilde{H}_{p^b(l-1)}^a(k) + (1-\alpha) \cdot \hat{H}_{p^b(l)}^a(k).
    \label{eq:smoothing_speechdep}
\end{equation}
The smoothed RTF $\tilde{H}_{p^b(l)}^a(k)$ is then used to simulate the own voice of talker $\mathrm{b}$ at the in-ear microphone:
\begin{equation}
    \hat{S}_\mathrm{i}^b(k,l) =  \tilde{H}_{p^b(l)}^a(k) \cdot Y_\mathrm{o}^b(k,l).
    \label{eq:simulation_speechdep}
\end{equation}
Similarly to the speech-independent model, a WOLA scheme is employed to obtain the time-domain signal $\hat{s}_i^b[n]$.
Figure~\ref{fig:signalflow_sim_phone_indiv} depicts the signal flow to simulate in-ear own voice signals for talker $b$ using the speech-dependent model for talker $a$.
Due to the phoneme recognition system for frame-wise phoneme-specific RTF selection, we expect that the proposed speech-dependent model is able to simulate in-ear signals more accurately than the speech-independent model, also for utterances not used 
during system identification.
In addition, it should be realized that unlike the speech-independent model, the speech-dependent model also accounts for speech pauses by modeling them as a separate phoneme.

\subsection{Talker-averaged models}
\label{sec:avgmodels}
Since individual models may generalize well to different talkers, we also consider talker-averaged speech-independent and speech-dependent models.      
In the \textit{system identification step}, talker-averaged models are obtained by considering all STFT frames of the recorded microphone signals of all utterances from all talkers except talker $b$ (leave-one-out-paradigm) for system identification. 
The RTFs of the speech-independent talker-averaged model are hence computed as
\begin{equation}
    \hat{H}^\mathrm{avg}(k) = \frac{\sum_{a \neq b}  \sum_l Y_\mathrm{i}^a(k,l) \cdot Y_\mathrm{o}^{a,*}(k,l)  }{ \sum_{a \neq b}  \sum_l |Y_\mathrm{o}^a(k,l)|^2},
\end{equation} 
while the RTFs of the speech-dependent talker-averaged model for phoneme $p^\prime$ are computed as
\begin{equation}
    \hat{H}_{p^\prime}^\mathrm{avg}(k) = \frac{\sum_{a \neq b} \sum_{p(l)=p^\prime} Y_\mathrm{i}^a(k,l) \cdot Y_\mathrm{o}^{a,*}(k,l) }{ \sum_{a \neq b}  \sum_{p(l)=p^\prime} |Y_\mathrm{o}^a(k,l)|^2}.
\end{equation}
The \textit{simulation step} for the talker-averaged models is similar as for  the individual models, where for the speech-independent model $\hat{H}^\text{avg}(k)$ is used instead of $\hat{H}^a(k)$ and for the speech-dependent model $\hat{H}_{p^\prime}^\mathrm{avg}(k)$ is used instead of $\hat{H}_{p^\prime}^a(k)$.

\subsection{Adaptive filtering-based model}
\label{sec:adaptmodel}
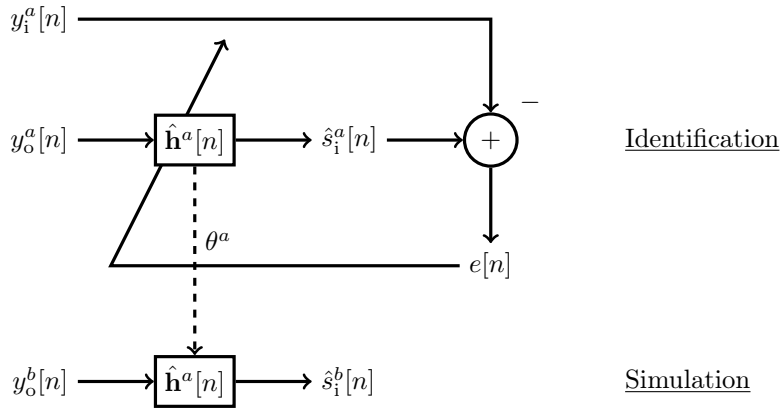
\begin{figure}
    \centering
        \begin{tikzpicture}
        \node (so) at (0,0) {$y_\mathrm{o}^a[n]$};
        \node[right=of so,draw,very thick,fill=white] (h) {$\hat{\mathbf{h}}^a[n]$};
        \node[right=of h] (sihat) {$\hat{s}_\mathrm{i}^a[n]$};
        \node[right=of sihat,label=above right:{$-$},circle,draw,very thick] (plus) {$+$};
        \node[above=of so] (si) {$y_\mathrm{i}^a[n]$};
        \node[below=of plus] (e) {$e[n]$};

        \node[below=2.5 of h,draw,very thick,fill=white] (h') {$\hat{\mathbf{h}}^a[n]$};
        \node[left=of h'] (yob) {$y_\mathrm{o}^b[n]$};
        \node[right=of h'] (sibhat) {$\hat{s}_\mathrm{i}^b[n]$};
        \draw[->, dashed, very thick] (h) -- (h') node[midway,above right] {$\theta^a$}; 
        \draw[->, very thick] (yob) -- (h');
        \draw[->, very thick] (h') -- (sibhat);

        \node[right=5 of h] (idlabel) {\underline{Identification}};
        \node[right=5 of h'] (simlabel) {\underline{Simulation}};
        
        \draw[->, very thick] (so) -- (h);
        \draw[->, very thick] (h) -- (sihat);
        \draw[->, very thick] (sihat) -- (plus);
        \draw[->, very thick] (si) -| (plus);
        \draw[->, very thick] (plus) -- (e);
        \begin{scope}[on background layer]
        \draw[->, very thick] (e) -- ++(-5,0) -- ++(1.5,3);
        \end{scope}
    \end{tikzpicture}
    \caption{The adaptive filtering scheme utilized for estimating in-ear speech signals. The filter coefficients are transferred from identification to simulation directly after each sample-wise adaptation step.} 
    \label{fig:nlms_diagram}
\end{figure}

As an alternative to the time-varying speech-dependent model in Section~\ref{sec:speechdepmodel}, in this section we consider a time-domain adaptive filter to model the time-varying transfer path between the microphone at the entrance of the occluded ear canal and the in-ear microphone. The signal flow is illustrated in Figure~\ref{fig:nlms_diagram}. 
In the \textit{system identification step}, 
the FIR filter $\hat{\mathbf{h}}^a[n]$ with $N$ coefficients is adapted based on recorded microphone signals of an utterance of talker $a$.
The adaptive filter aims at minimizing the error between the in-ear microphone signal $y_i^a[n]$ and the estimated in-ear own voice signal
\begin{equation}
    \hat{s}_\mathrm{i}^a[n] = \hat{H}^a(q,n) \cdot y_o^a[n] = \left(\hat{\mathbf{h}}^a[n]\right)^T \mathbf{y}_o^a[n], 
\end{equation}
with 
\begin{equation}
    \mathbf{y}_o^a[n] = \begin{bmatrix}
        y_o^a[n], & y_o^a[n-1], & \dots, & y_o^a[n-N+1]
    \end{bmatrix}^T.
\end{equation}
For adapting the filter the well-known normalized least mean squares (NLMS) algorithm is used~\cite{haykin_adaptive_1996}, i.e.~the filter coefficients are recursively updated as
\begin{equation}
\label{eq:adaptive_filter_update}
    \hat{\mathbf{h}}^a[n+1] = \hat{\mathbf{h}}^a[n] + \frac{\mu}{\epsilon + \left(\mathbf{y}_\mathrm{o}^a[n]\right)^T \mathbf{y}_\mathrm{o}^a[n]} \mathbf{y}_\mathrm{o}^a[n] \left( y_i^a[n] - \left(\hat{\mathbf{h}}^a[n]\right)^T \mathbf{y}_o^a[n] \right),
\end{equation}
where $\mu$ denotes the step size and $\epsilon$ is a small regularization constant.
The model parameters of the adaptive filtering-based model are
\begin{equation}
    \theta_\mathrm{adapt.}^a = \{\hat{\mathbf{h}}^a[n], n=1, \dots \}. 
\end{equation}
Since this model implicitly depends on a specific utterance, it should be noted that it is not possible to obtain a talker-averaged model by following a similar procedure as described in the previous section.

In the \textit{simulation step}, the simulated in-ear own voice signal of talker $b$ is computed as
\begin{equation}
    \hat{s}_\mathrm{i}^b[n] = \left(\hat{\mathbf{h}}^a[n]\right)^T \mathbf{y}_\mathrm{o}^b[n].
\end{equation}
In case of utterance mismatch (both for the same talker as well as for a different talker), the filter is applied to a different input signal than used during adaptation which likely results in estimation errors. 

\section{Experimental Evaluation}
\label{sec:evaluation}
In this section, the own voice transfer characteristic models discussed in Section~\ref{sec:models} are evaluated in terms of their accuracy in simulating in-ear own voice signals for different conditions.
In Section~\ref{sec:experiment_setup}, the data used in the evaluation and the experimental conditions are described.
In Section~\ref{sec:model_setup}, the simulation parameters are defined.
In Section~\ref{sec:example_spec_rtf}, examples of simulated in-ear own voice signals and estimated RTFs are presented for all considered RTF-based models.
In Sections~\ref{sec:results_sametalker}-\ref{sec:results_talkermismatch}, experimental results are presented and discussed for three conditions: matched condition (same talker, same utterance), utterance mismatch and talker mismatch. 

\subsection{Recording setup and experimental conditions}
\label{sec:experiment_setup}
For identifying and evaluating the own voice transfer characteristic models, we recorded a dataset of own voice speech from 18
native German talkers (5 female, 13 male), with approximately 25 to 30 minutes of recorded own voice signals per talker. 
The hearable device used for recording is the closed-vent variant of the one-size-fits-all Hearpiece~\cite{denk_one-size-fits-all_2019}.
The Hearpiece \textit{concha} microphone of the device was selected as the microphone at the outer face of the occluded ear canal. 
Talkers were excluded if insertion of the hearable was not possible, or if bad fittings with insufficient attenuation of external sounds were detected (by measuring a transfer function from an external loudspeaker between the concha and in-ear microphone).  
For each talker, 306 pre-determined sentences were recorded: 
The Marburg and Berlin sentences~\cite{simpson_kiel_1997}, each consisting of 100 sentences, 100 common everyday German sentences for language learners~\cite{neustein_100_2019}, and the German version of the well-known text \textit{The North Wind and the Sun}, consisting of 6 sentences. 
Recordings were conducted in a sound-proof listening booth using a Behringer UMC1820 audio interface.
Before the recordings started, informed consent was obtained from all talkers.
The recorded dataset is publicly available on Zenodo~\cite{ohlenbusch_german_2024}.
During system identification, model parameters were estimated on 150 sentences uttered by each talker. 
During simulation, in-ear own voice signals are generated from the recorded microphone signals at the outer face of the Hearpiece 
and evaluated per utterance.

%
Three different simulation conditions are investigated:
\begin{description}
    \item [Same talker, same utterance (matched condition)] 
    In this condition, the individual RTF-based models and the adaptive filtering-based model are evaluated exactly the same utterances of the same talker ($a=b$) as considered during model estimation. 
    For the adaptive filtering-based model, this means that the same signal $y_o^b[n]=y_o^a[n]$ is used during simulation as during identification (see Figure~\ref{fig:nlms_diagram}), such that the simulated in-ear signal $\hat{s}_i^b[n]$ is equal to the output of the adaptive filter $\hat{s}_i^a[n]$. 
    Talker-averaged models are not considered in this condition. 
    \item [Same talker, utterance mismatch] 
    In this condition, the individual RTF-based models and the adaptive filtering-based model are evaluated on speech of the same talker ($a=b$) as considered during model estimation. 
    In order to investigate the generalization ability of the models for the same talker, evaluation is performed on the 156 sentences not used to estimate the models. 
    For the adaptive filtering-based model, the length of the signals used during simulation and identification is matched, either by cutting or concatenating the signals used during model estimation with other signals from the same talker.
    Talker-averaged models are not considered in this condition.
    \item [Talker mismatch] 
    The generalization ability of models to unseen talkers is investigated by estimating speech of talker $b$ using models estimated on a different talker ($a \neq b$). 
    For each utterance, a random talker $a$ is assigned to talker $b$.
    In this condition, there is also an implicit utterance mismatch because the same sentence uttered by different talkers most likely has differences with respect to speed, frequency content, pronunciation and other speech attributes. 
    Talker-averaged models are considered in this condition only.
    For each talker $b$, a talker-averaged model is computed from utterances of the remaining 17 talkers.
    Evaluation is performed on the 156 sentences not used to estimate the models.
\end{description} %
In all three conditions, Log-Spectral Distance (LSD)~\cite{gray_distance_1976} and Mel-Cepstral Distance (MCD)~\cite{kubichek_mel-cepstral_1993} between the recorded in-ear signals $y_i^b[n]$ and the simulated in-ear signals $\hat{s}_i^b[n]$ are used as evaluation metrics. 
For both metrics, a lower value indicates a more accurate estimate. 
Since perceptual metrics such as PESQ~\cite{international_telecommunications_union_itu_itu-t_2001} were found not to correlate well with subjective ratings of body-conducted own voice signals~\cite{richard_comparison_2023}, such metrics are not considered in this study.

\subsection{Simulation parameters}
\label{sec:model_setup}
The experiments were carried out at a sampling frequency of 5\,kHz, since above 2.5\,kHz the in-ear microphone signals hardly contain any body-conducted speech for the considered hearable device.
Model-specific parameters were set empirically based on preliminary experiments.
For the RTF-based models, an STFT framework with a frame length of $K=128$ (corresponding to 25.6\,ms) and an overlap of 50\,\% was used, where a square-root Hann window was utilized both as analysis and synthesis window. 
For the speech-dependent models, a smoothing parameter of $\alpha=0.8$ was used in~\eqref{eq:smoothing_speechdep}, 
corresponding to an effective smoothing time of 64\,ms.
The used phoneme recognition system was trained on German speech and $P=62$ phoneme classes.
For the adaptive filtering-based model, the filter length was set to $N=128$, and a step size parameter $\mu=0.5$ and regularization constant $\epsilon=10^{-6}$ were used in~\eqref{eq:adaptive_filter_update}. The filter coefficients were initialized as zeroes.
For all methods, no voice activity detection was employed so that utterances may contain short pauses.

\subsection{Example spectrograms and RTFs}
\label{sec:example_spec_rtf}
\begin{figure}
    \centering
    \includegraphics{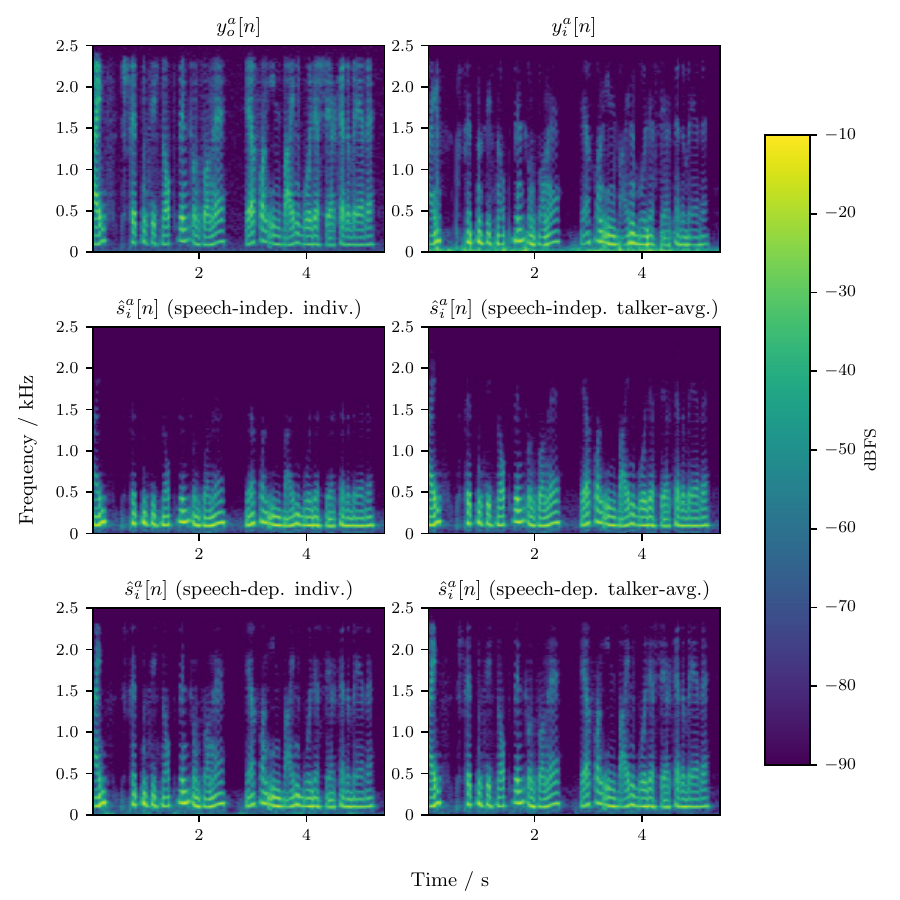} 
    \caption{Example spectrograms for the same talker, same utterance condition: recorded own voice signal of talker 2 at the entrance of the occluded ear canal
    (top left) and recorded in-ear own voice signal (top right) of talker 2, and the simulated in-ear own voice signals estimated by the speech-independent individual (middle left) and speech-independent talker-averaged (middle right), and the speech-dependent individual (bottom left) and speech-dependent talker-averaged (bottom right) models.} 
    \label{fig:specgram_examples}
\end{figure}

\begin{figure}
    \centering
    \includegraphics{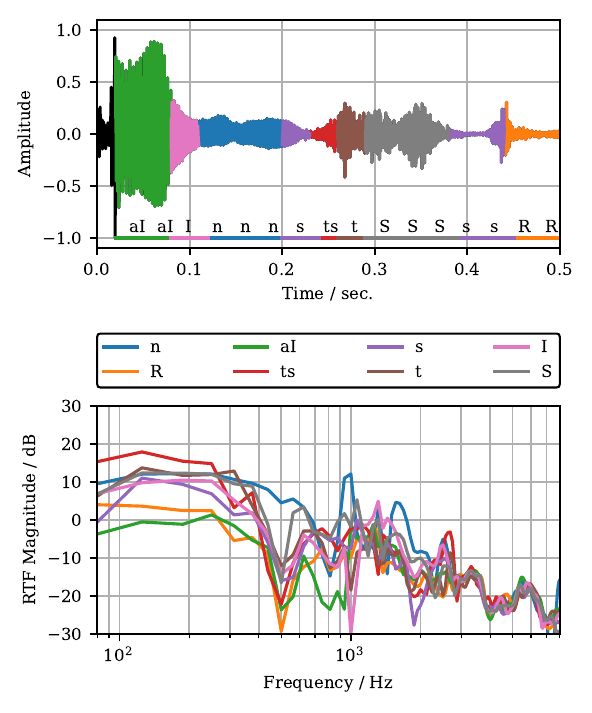} 
    \caption{
    Example own voice signal of talker 2 recorded at the entrance of the occluded ear canal with phoneme annotation (top) 
    and magnitude of phoneme-specific individual relative transfer functions (bottom)
    estimated on all utterances of this talker (speech-dependent individual model). 
    Only RTF magnitudes of phonemes appearing in the depicted utterance are shown.
    } 
    \label{fig:rtfs_timedomain_annotation_phones}
\end{figure}

For the RTF-based models, this section presents examples of simulated in-ear own voice signals, spectrograms and estimated RTFs.
For  the matched condition (same talker, same utterance), Figure~\ref{fig:specgram_examples} for a specific utterance (the beginning of \textit{The North Wind and the Sun}) of talker 2 (male).
The shown spectrograms are the spectrograms of the microphone signal at the entrance of the occluded ear canal and the in-ear microphone signal as well as the in-ear own voice signals simulated with the speech-independent models and the proposed speech-dependent models (individual and talker-average)\footnote{Audio examples corresponding to the spectrograms are available online at \url{https://m-ohlenbusch.github.io/own_voice_modeling_examples/}}. 
While it can be observed that the speech-independent models estimate the in-ear microphone signal rather well in the frequency region below 500\,Hz, they clearly underestimate own voice components for higher frequencies. 
On the other hand, the speech-dependent models are able to estimate the in-ear microphone signal more accurately at higher frequencies, although deviations are visible above 1\,kHz. 
The estimates of individual and talker-averaged models are very similar for both the speech-independent and speech-dependent models for this example.
It should be noted that the low-frequency body-produced noise in the in-ear microphone signal is not present in all simulated in-ear own voice signals. 
For the same utterance as in Figure~\ref{fig:specgram_examples}, 
Figure~\ref{fig:rtfs_timedomain_annotation_phones} depicts the time-domain own voice signal recorded at the entrance of the occluded ear canal with its phoneme annotation, and the magnitude of the phoneme-specific individual RTFs, estimated using~\eqref{eq:speechdep_indiv_estimation}.
Different from other experiments, these RTFs were estimated with a sampling frequency of 16\,kHz and an STFT size of $N=256$ to show the high-frequency region as well. 
It can be seen that for different phonemes, the RTFs differ a lot in the low-frequency region below 2.5\,kHz, while above 2.5\,kHz the RTFs are very similar.

\begin{figure}
    \centering
    \includegraphics{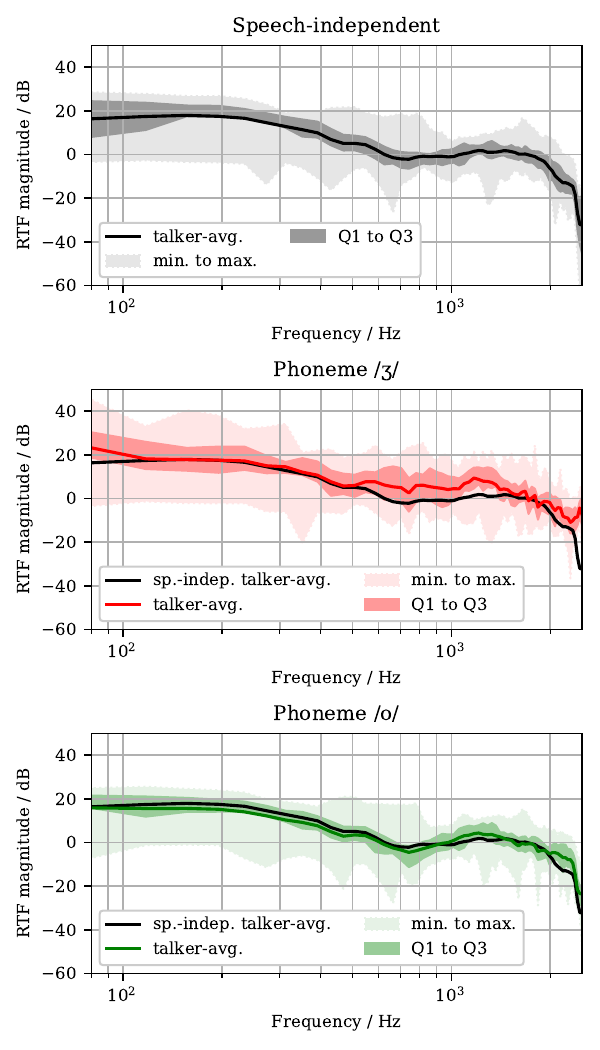} 
    \caption{Relative transfer functions estimated for the speeech-independent individual and talker-averaged models (top) and for two phonemes with the speech-dependent models (middle and bottom). Values between the quartiles Q1 and Q3 and between the minimum and maximum values of the individual models are indicated by shaded regions.
    Talker-averaged relative transfer functions over all talkers are shown as solid black lines.}
    \label{fig:rtf_schlaeuche}
\end{figure}

To compare the RTF-based models, Figure~\ref{fig:rtf_schlaeuche} depicts the estimated RTF magnitudes for the speech-independent models (top subplot) and the speech-dependent models for two selected phonemes (middle and bottom subplot), considering all talkers in the experiments.
The individual RTFs are represented by shaded regions and  the talker-averaged RTFs as solid lines.
Different from the talker-averaged RTFs used in the talker mismatch condition (leave-one-out-paradigm), averages here are computed over all 18 talkers.
For the speech-independent RTFs, it can be observed that for most talkers the low frequency region below approximately 600\,Hz is amplified at the in-ear microphone relative to the microphone at the entrance of the occluded ear canal, whereas the frequency region above approximately 1.5\,kHz is attenuated.
While half of the estimated RTFs (i.e., between the quartiles Q1 and Q3) are very similar in magnitude, for some talkers there appear to be larger deviations from the talker-averaged RTF magnitude. 
For the phoneme-specific RTFs shown in the middle and lower subplot, similar tendencies in terms of inter-individual variance can be observed. 
However, it can be observed that the phoneme-specific talker-averaged RTFs differ from the speech-independent talker-averaged RTFs. 
In particular, for the phoneme \textipa{/Z/} the magnitude is considerably higher than the magnitude of the speech-independent talker-averaged RTF in the frequency region between 500\,Hz to 1.5\,kHz and above 2\,kHz for the majority of talkers. 
In contrast, for the phoneme \textipa{/o/} the RTF magnitudes are lower than the magnitude of the speech-independent talker-averaged RTF especially in the low frequency region.

\subsection{Same talker, same utterance} 
\label{sec:results_sametalker}
\begin{figure}
    \centering
    \begin{subfigure}[b]{0.49\textwidth}
        \includegraphics{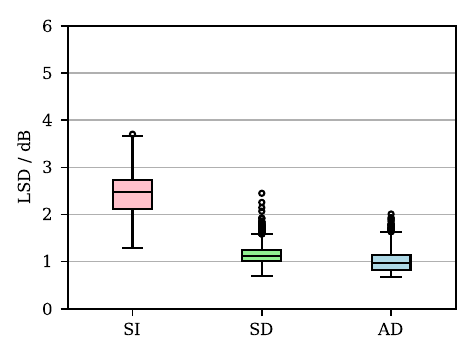}
        \caption{Log-Spectral Distance.} 
        \label{fig:lsd_ST_SU}
    \end{subfigure}
    \hfill
    \begin{subfigure}[b]{0.49\textwidth}
        \includegraphics{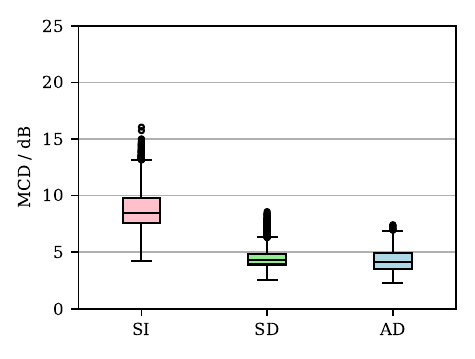}
        \caption{Mel-Cepstral Distance.}
        \label{fig:mcd_ST_SU}
    \end{subfigure}
    \caption{Results for the \textit{same talker, same utterance} condition with speech-independent (SI), speech-dependent (SD) and adaptive filtering-based (AD) models.}
    \label{fig:results_ST_SU}
\end{figure}

For the matched condition (same talker, same utterance), Figure~\ref{fig:results_ST_SU} shows the LSD and MCD scores between the recorded in-ear signals and the simulated in-ear signals for the speech-independent and speech-dependent individual RTF-based models and the adaptive filtering-based model.
It can be observed that both metrics are much lower for the speech-dependent individual model and the adaptive filtering-based model than for the speech-independent individual model.
These results demonstrate that in-ear own voice signals can be simulated more accurately when time-varying or speech-dependent transfer characteristics are accounted for. 
In addition, the speech-dependent individual model performs nearly as well as the adaptive filtering-based model, where it should be realized that for the matched condition the (utterance-specific) adaptive filter can be considered as the optimal time-varying filter.
This indicates that the proposed phoneme-specific RTF-based model is able to accurately model time-varying behavior of own voice transfer characteristics. 
It can be noted that even in the matched condition none of the considered methods is able to perfectly simulate the recorded in-ear own voice signals. 
This can be explained by the fact that the considered methods are not able to account for body-produced noise (see Figure~\ref{fig:specgram_examples}) and possible non-linear effects, which are however assumed to be small.

\subsection{Same talker, utterance mismatch}
\label{sec:results_utterancemismatch}
\begin{figure}
    \centering
    \begin{subfigure}[b]{0.49\textwidth}
        \includegraphics{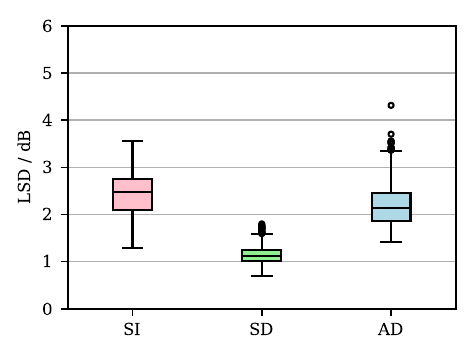}
        \caption{Log-Spectral Distance.}
        \label{fig:lsd_ST_DU}
    \end{subfigure}
    \hfill
    \begin{subfigure}[b]{0.49\textwidth}
        \includegraphics{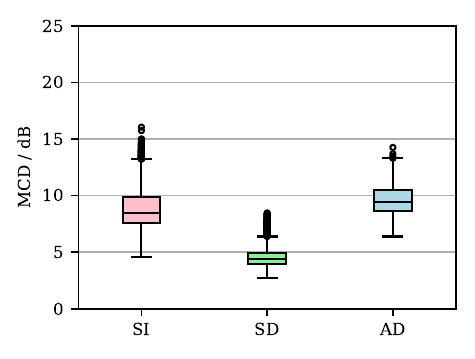}
        \caption{Mel-Cepstral Distance.}
        \label{fig:mcd_ST_DU}      
    \end{subfigure}
    \caption{Results for the \textit{same talker, utterance mismatch} condition with speech-independent (SI), speech-dependent (SD) and adaptive filtering-based (AD) models.}
    \label{fig:results_ST_DU}
\end{figure}

For the same models as in the previous section, Figure~\ref{fig:results_ST_DU} shows the LSD and MCD score for the utterance mismatch condition (same talker, utterance mismatch). 
The results for the speech-dependent and speech-independent individual models are very similar matched condition (see Figure~\ref{fig:results_ST_SU}), indicating that both models generalize well to other utterances of the same talker. 
For the adaptive filtering-based model, on the other hand, the LSD and MCD scores are much larger than for the matched condition, showing that the utterance-specific adaptive filtering-based method (expectedly) does not generalize well to other utterances.

\subsection{Talker mismatch} 
\label{sec:results_talkermismatch}
\begin{figure}
    \centering
    \begin{subfigure}[b]{0.49\textwidth}
        \includegraphics{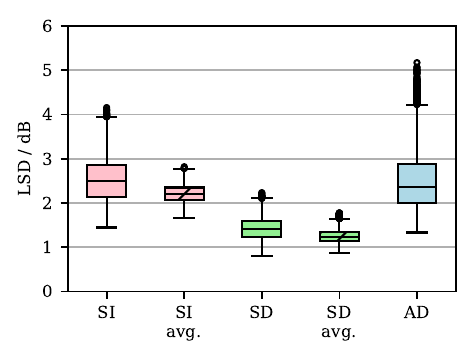}
        \caption{Log-Spectral Distance.} 
        \label{fig:lsd_DT_DU}
    \end{subfigure}
    \hfill
    \begin{subfigure}[b]{0.49\textwidth}
        \includegraphics{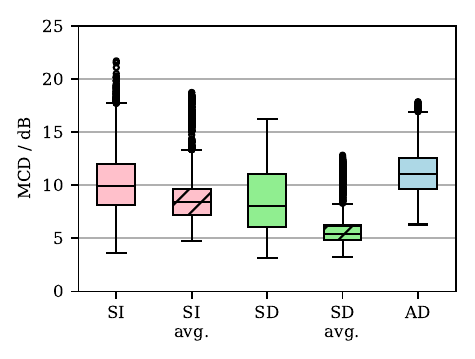}
        \caption{Mel-Cepstral Distance.}
        \label{fig:mcd_DT_DU}      
    \end{subfigure}
    \caption{Results for the \textit{talker mismatch} condition with speech-independent (SI), speech-dependent (SD) and adaptive filtering-based (AD) models, using individual and talker-averaged (avg.) versions.}
    \label{fig:results_DT_DU}
\end{figure}

For the talker mismatch condition, Figure~\ref{fig:results_DT_DU} shows the LSD and MCD scores for the speech-independent and speech-dependent models (both individual as well as talker-averaged) and the adaptive filtering-based model.
It can be clearly observed that the speech-dependent models outperform the speech-independent models and the adaptive filtering-based model, where the best performance in terms of both metrics is achieved by the speech-dependent talker-averaged model. 
This indicates that the speech-dependent talker-averaged model has the best generalization ability to unseen talkers.
Comparing the results in Figure~\ref{fig:results_ST_DU} and Figure~\ref{fig:results_DT_DU}, it can be observed that the LSD and MCD scores of the speech-dependent individual model are larger under talker mismatch. Especially the large variance of the MCD score is noticeable.
Since this effect does not occur in the other conditions, it is likely a consequence of talker mismatch. 

\section{Conclusion}
\label{sec:conclusion}
In this paper, speech-dependent models of own voice transfer characteristics in hearables have been proposed.
The models can be utilized to estimate own voice signals at an in-ear microphone. 
In particular, the proposed models take into account time-varying speech-dependent behavior and inter-individual differences between talkers.
To estimate in-ear own voice signals from broadband speech using the proposed speech-dependent models, phoneme-specific RTFs are used.
The influence of utterance and talker mismatch on the estimation accuracy of in-ear own voice signals has been investigated in an experimental evaluation.
Results show that using a speech-dependent model is beneficial compared to using a speech-independent model. 
Although the adaptive filtering-based approach is able to model the speech-dependency of the own voice transfer characteristics well in the matched condition, it completely fails when considering utterance and talker mismatch.
However, the proposed individual speech-dependent models are able to generalize to different utterances of the same talker.
Talker-averaged models were shown to generalize better to different talkers than individual models.
Future work will investigate the usage of the proposed models for simulating in-ear signals to train own voice reconstruction algorithms based on supervised learning.

\section*{Conflict of interest}
The authors declare no conflict of interest.

\section*{Acknowledgments}
The Oldenburg Branch for Hearing, Speech and Audio Technology HSA is
funded in the program \frqq Vorab\flqq~by the Lower Saxony Ministry of Science and
Culture (MWK) and the Volkswagen Foundation for its further development.
This work was partly funded by the German Ministry of Science and Education BMBF FK 16SV8811 
and the Deutsche Forschungsgemeinschaft (DFG, German Research Foundation) - Project ID 352015383 - SFB 1330 C1.
The authors wish to thank the talkers for their participation in the recordings.

\section*{Data Availability Statement}
The research data associated with this article are available in Zenodo, under the reference \url{https://zenodo.org/doi/10.5281/zenodo.10844598}.

\printbibliography

\end{document}